\documentclass[12pt,a4paper]{article}
\usepackage{amsmath,amssymb}
\usepackage{epsfig,graphicx}
\newcommand{\be}{\begin{equation}}
\newcommand{\ee}{\end{equation}}
\newcommand{\vv}{\upsilon}

\title{Dark energy: myths and reality\footnote{Published: Physics-Uspekhi {\bf 51}(3), pp 283-289 (2008).}}
\author{V.N. Lukash$^\dagger$, V.A. Rubakov$^\ddagger$\\
       $^\dagger${\small Astro Space Centre of P.N. Lebedev Physical
       Institute},\\
       {\small lukash@asc.rssi.ru};\\
     $^\ddagger${\small Institute for Nuclear Research},\\{\small rubakov@ms2.inr.ac.ru} }

\begin{document}
\maketitle

\begin{abstract}

We discuss the questions related to dark energy in the Universe.
We note that in spite of the effect of dark energy, large-scale
structure is still being generated in the Universe and this will
continue for about ten billion years. We also comment on some
statements in the paper ``Dark energy and universal
antigravitation'' by A.D. Chernin [4].

\end{abstract}

\section{Introduction}

The emergence of the idea that the entire visible Universe is
permeated by a weakly interacting substance known as dark energy
was the number one sensation in physics at the turn of the century
and came as a complete surprise to most scientists, particularly
those studying topics related both to cosmology and to particle
physics. This is because the known energy scales of fundamental
interactions are of the order of 1 GeV for the strong interaction
and 100 GeV and $10^{19}$ GeV for the weak and gravitational
interactions, respectively. Thus, there was no
reason\footnote{Apart from some arguments based on the anthropic
principle, see [1-3].} to assume that a new energy scale much
smaller than the above-mentioned ones exists in nature. But it
turned out that dark energy is characterized by the
scale\footnote{We follow the notation used in Chernin's article
[4]: the subscripts V, D, and B respectively stand for dark
energy, dark matter, and baryons.} $E_{\rm V} \sim 10^{-3}$ eV
defined by the relation $\rho _{\rm V} = E_{\rm V}^{\,4}$, where
$\rho _{\rm V}$ is the dark energy density.

Moreover, in the present-day Universe, the following equality is
valid within an order of magnitude:
$$
\rho _{\rm V} \approx \rho _{\rm D} \approx \rho _{\rm B}\,,
\eqno(1)
$$
where $\rho _{\rm D}$ and $\rho _{\rm B}$ are mass densities of
dark matter and baryons (protons and nuclei). And again, there is
no clear a priori reasons for this equality. We point out that the
approximate relation $\rho _{\rm D} \approx \rho _{\rm B}$ has
been valid at each instant of cosmological evolution since the
baryon asymmetry emerged and dark matter was generated, because
$\rho _{\rm D}$ and $\rho _{\rm B}$ decrease with the expansion of
the Universe at the same quite high rate. On the other hand, $\rho
_{\rm V}$ does not depend, or barely depends, on time; hence, it
is today, i.e., after the structure has appeared and stars have
been formed, that the first equality in (1) is satisfied. It is
certainly not easy to admit that relation (1) holds merely
accidentally.

Because the properties of dark energy are very interesting and the
problem itself is fundamental, it is important to understand what
kind of data made scientists believe that dark energy does exist.
This knowledge is necessary when we try to find explanations,
which may seem exotic today, of why the expansion of the Universe
is accelerating and when we choose key experiments to verify
various hypotheses. We mention one such attempt below. Attempts to
explain approximate relation (1) also deserve attention.

These issues are primarily considered in A.D. Chernin's paper [4].
However, we believe that that paper somewhat mythologizes dark
energy because on the one hand, some crucial results are hardly
mentioned there, and on the other hand, some issues discussed in
that article have nothing to do with dark energy. Besides, the
attempt to explain relation (1) using some `internal symmetry in
cosmology' is, to put it mildly, highly disputable.

In this paper, we try to separate the dark energy mythology from
the real state of affairs.

\section{Structure argument and supernovae}

Type-Ia supernovae are often used (Chernin also writes about them)
as the main observational argument confirming the existence of
dark energy. But there are quite a number of other arguments, at
least equally serious, based on combinations of cosmological data.
Some of them were known before the observational data on type-Ia
supernovae appeared, which had made several cosmologists [5-9]
(see [10] for a review of the earlier papers) insist on the
existence of dark energy in nature even before the first results
on supernovae were available.

One of the independent arguments is as follows. By the mid-1990s,
the analysis of galactic catalogues aimed at revealing the
distribution of matter in space, the use of various methods to
determine the mass of clustered (clumped) matter\footnote{
Measurements of peculiar velocities of galaxies in clusters and
superclusters, gravitational lensing by clusters, measurements of
galaxies' rotation curves, determination of the mass--luminosity
relation, measurements of X-ray clusters' temperatures, etc.} and
measurements of the cosmic microwave background and the Hubble
parameter had already led to the conclusion that the total mass
density of nonrelativistic matter, which constitutes the
inhomogeneous structure of our Universe, such as galaxies and
their formations (groups, clusters, filaments, walls,
superclusters, voids), is not greater than 30 percent of the
critical density $\rho _{\rm c}$:
$$
\Omega _{\rm M} \equiv {\rho _{\rm M}\over \rho _{\rm c}} \,\le
\,0.3 \,,        \eqno(2)
$$
where
$$
\rho _{\rm M} = \rho _{\rm D} + \rho _{\rm B} \,, \qquad \rho
_{\rm c} = {3 H_0^{\,2}\over 8 \pi G} \simeq 10^{-29}\;{\rm
g\;cm}^{-3}\,,
$$
and $H_0 \simeq 70\,{\rm km}\,{\rm s}^{-1}{\rm Mpc}^{-1}$ is the
Hubble parameter. Result (2) is one of the most important facts in
modern cosmology. For a long time, it has been interpreted as
evidence of the Universe having nonzero curvature. Indeed, if dark
energy is not taken into account, then the Friedmann equation for
the open cosmological model, written for the present epoch,
reduces to the relation
$$
\rho _{\rm c} = \rho _{\rm M} + {3\over 8 \pi G R_\kappa ^2} \,,
$$
where $R_\kappa$ is the present curvature radius of space.
According to (2), the curvature (the second term in the right-hand
side) dominates, giving not less than $0.7 \rho _{\rm c}$.

But this interpretation faced difficulties. First, from the
theoretical standpoint, a distinctly nonzero spatial curvature is
almost incompatible with the inflationary Universe paradigm
because inflationary models without fine tuning result in
extremely small values of the spatial curvature $R^{\,-2}_\kappa$.
Second, the present age of the Universe in the open model is
around 11 billion years, whereas estimations of the age of the
oldest objects in the Universe (for example, globular clusters)
have yielded greater values of 12 to 14 billion years. There were
also other arguments against the open model with a large spatial
curvature.

If the spatial curvature is zero, result (2) suggests that not
less than 70\% of the energy density in the modern Universe is due
to matter of a type that cannot be perturbed by gravitational
fields of the structures and remains unclumped (unclustered) in
the course of cosmological evolution. This implies that the
effective pressure of the matter is negative\footnote{We note that
the condition of dominance over nonrelativistic matter excludes
relativistic particles because their energy density decreases at a
higher rate than the nonrelativistic matter density.} and its
absolute value is sufficiently large, i.e., $p \approx -\rho$.
Hence, this is dark energy.

The model with spatial curvature was finally discarded based on
measurements of the cosmic microwave background anisotropy, or, to
be more precise, the determination of the first peak position in
the angular spectrum of the anisotropy, this peak being most
sensitive to the value of the spatial curvature. Thanks to these
measurements, it was already clear in 1999--2000 that the
three-dimensional space is Euclidian to a high precision (i.e.,
$R_\kappa ^{\,-1}$ is close to zero). Here, a key role was played
by the balloon experiments BOOMERANG (Balloon Observations of
Millimetric Extragalactic Radiation and Geophysics) and MAXIMA
(Millimeter Wave Anisotropy Experiment Imaging Array) [11--14].
Later, the WMAP (Wilkinson Microwave Anisotropy Probe) experiment
and others confirmed this result. Thus, the total energy density
of all sorts of matter must indeed coincide with the critical
density and, hence, dark energy does exist in nature.

The structure argument based on the measurements of the microwave
radiation anisotropy and polarization, combined with data on the
large-scale structure of the Universe, is presently a clear
evidence for the existence of dark matter. We also note the
integrated Sachs--Wolfe effect, which has recently been confirmed
in observations. In the future, this effect should become one of
the most precise methods to measure the properties of dark energy
[15].

To illustrate the importance of the combination of all
cosmological data on dark energy, we mention, for example, an
attempt to explain the type-Ia supernova observational data
alternatively (see review [16] and a recent discussion in [17])
based on an assumption that matter density in our part of the
Universe is significantly lower than the average value. Analysis
shows [17] that this model may be consistent not only with the
supernova data but also with those on microwave radiation. But
whether the model will fit the results on large-scale structure
and other cosmological data is highly questionable.

\section{Hubble flows and their distortion}

Among the independent arguments supporting the existence of dark
energy, Chernin considers the measurements of local cold flow (the
author uses the term `Hubble flow,' which is not quite accurate),
which he discusses in great detail. Unfortunately, here we face a
mythologization of dark energy.

The main theses in paper [4] and some previous articles by Chernin
et al. are based on the statement that Hubble's law manifests
itself even at cosmologically small distances, which is explained
by antigravitation. For example, on pages 278 and 279, we read,
``...antigravitation is actually capable of driving galaxies'
motion almost in the entire range of cosmological distances, both
at the global, `genuinely' cosmological scales and at scales of
just a few megaparsecs,'' ``...antigravitation also dominates in
our nearest galactic environment at the distance of just 1-2 Mpc
from the Milky Way,'' ``...it is the dark energy... that actually
lies behind Hubble's discovery and makes sense of it for
cosmology,'' ``...dark energy can be... measured in every place
where a regular outflaw of galaxies is observed.'' But we show in
this section that, as a matter of fact, dark energy has not yet
influenced the local velocity distribution to the full extent, and
the expansion in accordance with the Hubble law starting from the
scale of several megaparsecs is excluded. What does define the
local flow properties is the profile of the spatial density
perturbation spectrum.

The initial Hubble flows existed throughout the entire Universe.
The flows in different regions were destroyed at different times,
in a manner that directly depended on the forming structure.

As is known, the structure of the Universe has resulted from
gravitational amplification of density perturbations whose initial
amplitudes were about $10^{-5}$ for wavelengths equal to the
Hubble size at that time. The perturbations were growing faster
for short waves. As a result, the nonlinearity scale at which the
Hubble flows are completely destroyed (the dark matter and baryon
perturbations $\delta _{\rm M} \equiv  \delta \rho _{\rm M} / \rho
_{\rm M} \sim 1$), was increasing with time. In the present-day
Universe, the average value of this scale is around 15 Mpc,
varying, however, between different regions of the Universe. For
example, it is smaller at a far distance from galaxy clusters
(which are the most massive gravitationally bound formations). In
particular, the nonlinearity scale in our local region is around
2~Mpc (the size of the Local Group of galaxies).

In quasilinear regions, where the density perturbations are still
not high ($\delta _{\rm M} < 1$), galaxies continue outflowing in
accordance with the initial conditions. But the Hubble flows in
such regions are also distorted. In the future, in dozens of
billions of years, peculiar velocities will fade out because of
the dynamic influence of dark energy, and the motion of galaxies
will obey the Hubble law\footnote{Here, we are talking only about
galaxies that are located in the quasilinear regions of space
($\delta _{\rm M} < 1$). In gravitationally bound systems ($\delta
_{\rm M} > 1$), galaxies held by the gravitational field of these
formations do not experience outflow.} again, as in the early
Universe.

This is our main disagreement with the thesis in [4]. Chernin
believes that outside the gravitationally bound regions peculiar
velocities of the galaxies have faded out owing to the dynamic
influence of dark energy, and the motion obeys the Hubble law. In
this section, we show that such recovery of the Hubble flows is
only possible in the distant future (if dark energy has vacuum
properties), as opposed to today, when the Universe is
experiencing peculiar velocities that have maximum values over its
history and are caused by inhomogeneities of matter density.

\subsection{Inhomogeneous Universe}

At the quasilinear stage, our Universe is described [18] by the
generalized Friedmann equation\footnote{At this stage, we neglect
the radiation density. The first two terms in the right-hand side
of (3) respectively describe nonrelativistic matter (dark matter
and baryons) and dark energy modeled by the cosmological constant.
The dot over a variable denotes the partial derivative with
respect to time t. The modern value of the cosmological scale
factor is set to $a_0 = 1$ [see (5)].}
$$
\biggl ({\dot b\over H_{\rm V}}\biggr )^2 = {c\over b} + b^{\,2}
-\kappa \equiv f^{\;2}(b) -\kappa ({\bf x})\,,       \eqno(3)
$$
where $(t,{\bf x})$ are Lagrangian coordinates comoving with
matter (the matter 4-velocity is $u_\alpha =t_{,\alpha }$), $b=
b(t,{\bf x})$ is the volume expansion scale factor [the comoving
matter density is equal to $\rho_{\rm M}=3c H_{\rm V}^{\,2}/(8\pi
Gb^{\,3})$], $H_{\rm V}=H_0\sqrt {\Omega _{\rm V}}\simeq  2\times
10^{-4}\,{\rm Mpc}^{-1}$ is the Hubble parameter of dark energy,
and
$$
c\equiv {\Omega _{\rm M}\over \Omega _{\rm V}} = {\Omega _{\rm
M}\over 1-\Omega _{\rm M}}\simeq 0.39 \,.
$$
The function
$$
f\,(b)\equiv \biggl ({c\over b}+b^{\,2}\biggr )^{1/2}\ge 1
$$
has a minimum $f_{\min} \simeq 1$ at $b_{\min}^{-1}\simeq 1.7$.

An arbitrary small function $\kappa = \kappa ({\bf x})$ of spatial
coordinates describes the local spatial curvature. We are
interested in the spatial regions where the right-hand
side\footnote{Condition (4) includes both superclusters (the
regions where $\kappa >0$) and cosmological voids ($\kappa <0$).}
of Eqn'(3) is positive:
$$
\kappa ({\bf x}) <1\,. \eqno(4)
$$
If this condition is satisfied, the matter density decreases with
time monotonically.

When $\kappa =0$, the volume and background scale factors are
equal (although the expansion anisotropy remains large; see
Section 3.2),
$$
b =a(t)\equiv {1\over 1+z}\,,\qquad H\equiv {\dot a\over a}=H_{\rm
V}\,{\,f\,(a)\over a}\,,       \eqno(5)
$$
where $\,f=f\,(a)$ is the growth rate factor of the Hubble
velocity (${\bf V}_{\rm H}=f\,H_{\rm V}{\bf x}$).

In the general case, in the linear order in $\kappa$, we obtain
$$
b=a\biggl (1-{1\over 3}\,g\kappa \biggr )\,,\qquad \delta _{\rm
M}=g\kappa \,,  \eqno(6)
$$
$$
H_{\rm eff}\equiv {\dot b\over b} =H\biggl (1- {1\over 3}\,h
\kappa \biggr )\,,\qquad h\equiv {\vv \over f} = {\dot {g}\over
H}\,,
$$
where $\delta _{\rm M}\equiv \delta \rho _{\rm M}/\rho _{\rm M}$
is the comoving density perturbation, $H_{\rm eff}=H_{\rm eff}
(t,{\bf x})$ is the effective Hubble function, and $g= g(a)$ and
$\vv =\vv (a)$ are the respective growth factors of the density
perturbations and the matter peculiar velocity [see also (11)],
$$
g(a)={1\over c} \biggl (a-H\int _0^{\,a} \,{{\rm d}a\over H}\biggr
)\,,\qquad \vv (a)={3H_{\rm V}\over 2a^{\,2}}\int _0^{\,a} \,{{\rm
d}a\over H}\,. \eqno(8)
$$

Equations (3)--(8) describe quasi-Hubble flows with the effective
Hubble parameter $H_{\rm eff}$ depending on the observer's
location. Figure 1 shows the functions $g(a)$ and $\vv (a)$. In
the present era, the function $\vv$  is in its wide maximum,
indicating the period of the most intensive structure formation.
The position of the maximum of $\vv (a)$ corresponds to $z\simeq
0.2$, the level of 90 percent of the maximum value is reached at
$a\simeq 0.5$ and 1.4, and the half-maximum is at $a\simeq 0.1$
and 4. Therefore, the present era is an era of maximum peculiar
velocities, and it will continue to last for a cosmological time.
The function $\vv$  will have decreased to only half its current
value by the time the Universe is 35 billion years old. And only
then will it be possible to talk about the era of faded-out
peculiar velocities in every space region where $\kappa <1$.

\begin{figure}[t]
\begin{center}
\includegraphics{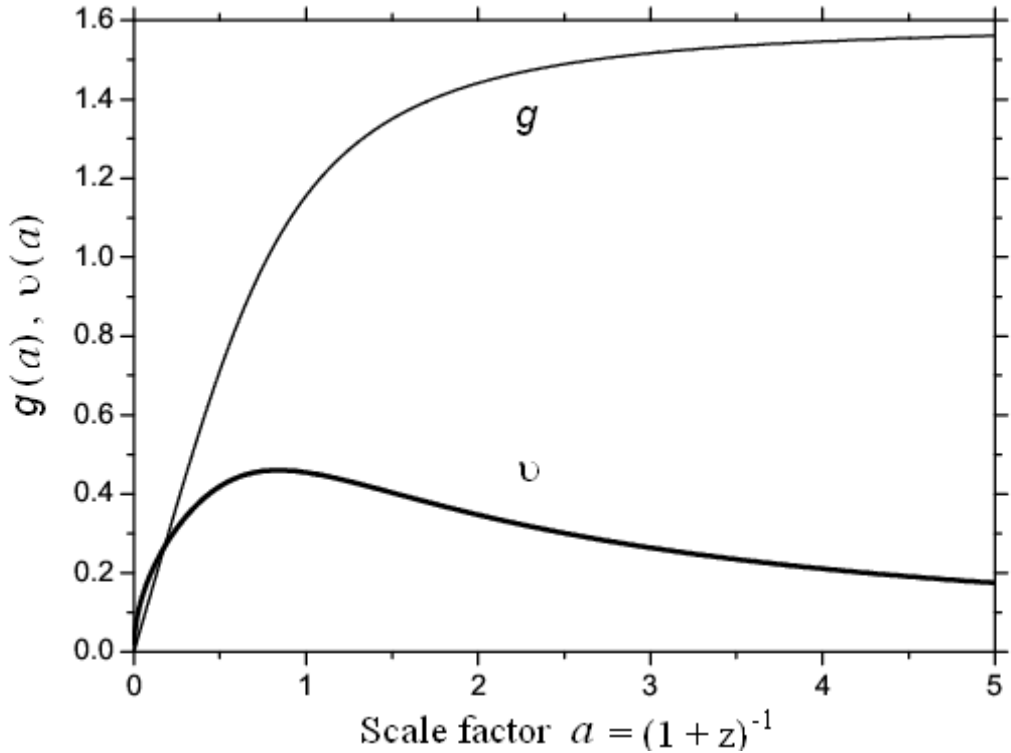}
\end{center}
\caption{The functions of the density perturbation growth rate
$g(a)$ and of the matter peculiar velocity $\vv (a)$.}
\end{figure}

Figure 2 displays the function $h=h(a)$ describing the deviation
of the local Hubble parameter from the background one. The
function has its maximum at $z\simeq 0.4$, and the interval $h
> 0.5 \, h_{\max}$ spans the range $a\in (0.1, 1.8)$, which corresponds
to the age 0.6 to 22 billion years. We can learn from Fig. 2 that
our Universe is at the stage of maximum distortion of Hubble's
expansion, and the Hubble flows are to be recovered only in some
10 billion years.

\begin{figure}[t]
\begin{center}
\includegraphics{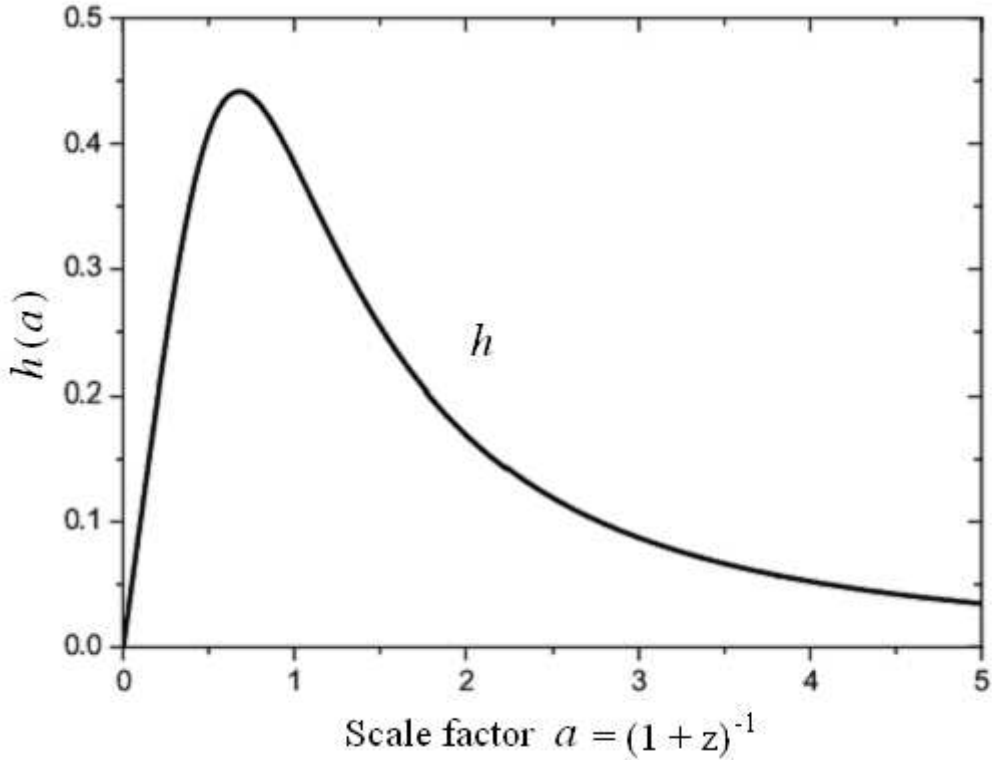}
\end{center}
\caption{The function $h(a)$ describing the deviation of the local
Hubble parameter $H_{\rm eff}$ from the background one.}
\end{figure}

To summarize, we can conclude that the large-scale structure
formation in the Universe occurs during the period that spreads
from 1 to 20 billion years since the Big Bang. The stage of the
suppression of Hubble flow inhomogeneities due to dark energy's
gravitational influence has not come yet. That is why one of the
key theses in [4, p. 279] --- that `the dynamic effect of dark
energy naturally explains the two astronomical facts that have
seemed mysterious up to now: (1) regularity of the expansion flow
inside the uniformity cell and (2) the same expansion rate at
local and global scales' --- is incorrect.

We now consider local matter flows and their properties in more
detail.

\subsection{Anisotropy of cold flows}

To describe the gauge-invariant field of peculiar velocities, we
pass to Euler (quasi-Friedmannian) coordinates $y^{\,\alpha } =
(\tau , {\bf y}$). In these coordinates, the gravitational field
is locally isotropic at any spatial point in the linear order in
$\kappa$. The required transformation is given by
$$
{\bf y}={\bf x}+g{\bf S}\,,\qquad \tau =t- a\vv H_{\rm V}\bar q\,,
\eqno (9)
$$
$$
S_i=-\bar q_{,\,i}\,,\qquad q= {3\over 2}\,H_{\rm V}^{\,2}\bar
q\,,
$$
where ${\bf S}={\bf S}({\bf x})$ is the vector of a medium element
displacement from the unperturbed position\footnote{We recall that
the {\bf x} coordinate does not change with time along the medium
element trajectory and tends to ${\bf y}$ as $t\rightarrow 0$.}
and $q= q({\bf x})$ is a time-independent dimensionless
displacement potential [18]. Comparison with (6) yields
$$
\kappa =- {\rm div} \, {\bf S} = \Delta \bar q\,.
$$
The matter particle displacement from the unperturbed Hubble
trajectories is monotonically increasing with time and today
amounts to the value $\sigma_S \simeq 15$ Mpc. If dark energy has
vacuum properties, the displacement will tend to its asymptotic
value of about 25 Mpc in the future.

Transformation (9) yields the following form of the metric
interval in the quasi-Friedmanninian and Lagrangian frames:
$$
{\rm d} s^{\,2} = (1+2\Phi ) {\rm d} \tau^{2} - \tilde a^{2} {\rm
d} {\bf y}^{2}
$$
$$
= {\rm d} t^{2} - \tilde a^{2} (\delta _{ik} - 2 g \bar q_{,ik})
{\rm d} x^{i} {\rm d} x^{k}\,, \eqno(10)
$$
where $\tilde a\equiv a(t)(1-q)$ is a 4-scalar invariant under
coordinate transformations and $\Phi =(cg/a)q$ is the
gravitational potential of density perturbations. The function
$b(t,{\bf x})$ is directly proportional to the trace of the
spatial part of the Lagrangian metric tensor [compare (10) and
(6)].

Equation (10) gives the physical Euler coordinate of the medium
element, ${\bf r}=\tilde a{\bf y}$. Differentiating ${\bf r}$ with
respect to the proper time yields the following expression for the
matter peculiar velocity:
$$
{\bf v}\equiv \dot {\bf r} - H{\bf r} =\vv \,H_{\rm V}\,{\bf S}\,.
\eqno(11)
$$
Expression (11) coincides with the definition of the 3-velocity as
the spatial component of the matter 4-velocity in the
quasi-Friedmannian reference frame:
$$
\vv _i=-{\partial t\over a\partial y^{\,i}}=-\vv \,H_{\rm V}\,\bar
q_{,\,i}\,.
$$
Therefore, the value $\vv$  appearing in (7) is indeed the
peculiar velocity growth rate.

We now consider the local Hubble flows. In regions (4) of the
inhomogeneous Universe, the Hubble flows are described by the
tensor field $H_{ik}=H_{ik}(t,{\bf x})$ generalizing the function
$H(t)$ in the Friedmann model [see (13)]. At a fixed instant of
time $t$, Eqns (9) give the coordinate distance between two close
medium points:
$$
\delta y_i=(\delta _{ik} - g \bar q_{,\,ik})\delta x^{\,k}\,.
\eqno(12)
$$
Differentiating the physical distance $\delta {\bf r}=\tilde
a\,\delta {\bf y}$ with respect to time yields the field of paired
velocities:
$$
\delta V_i\equiv {\partial \over
\partial t} (\delta r_i) =H_{ik}\,\delta r^{\,k}\,, \eqno(13)
$$
$$
H_{ik}=H\delta _{ik} -\dot g\bar {q}_{,\,ik}=H (\delta _{ik}- h
\bar q_{,\,ik})\,.   \eqno(14)
$$
We can see that the trace of (14) corresponds to the volume Hubble
parameter $H_{\rm eff}=(1/3)H_{i\,i}$, but the tensor $H_{ik}$
itself is highly anisotropic. The local expansion anisotropy
(variations in the projections of $H_{ik}$ on directions radiating
from a given point ${\bf x}$) is of the same order as the
deviations of $H_{\rm eff}$ from the true Hubble parameter $H$.
For example, in the Local Group, at a distance more than 2 Mpc
from its barycenter, the expected anisotropy of the quasi-Hubble
outflaw of galaxies can amount to 30\%.

The field $H_{ik}$ describes regular (cold) matter flows. It is
worth saying that formula (13) is valid in the limit of small
distances between galaxies, i.e., distances smaller than the
correlation scale of the two-point correlation function of the
displacement vector. The correlation radius varies from 15 to 40
Mpc between projections of this vector with respect to the
direction $\delta {\bf y}$. As the distance increases, random
deviations from law (13) increase. This is due to the cosmological
velocity perturbation spectrum, whose amplitude decreases with a
decrease in the wavelength for $k > 0.03 \,{\rm Mpc}^{-1}$, and
hence the random deviations from average velocities (13) increase
as the wavelength $k^{-1}$ increases. Just to give an example, at
the distance 3 Mpc from the Local Group barycenter, the deviations
are around $30-40\,{\rm km}\, {\rm s}^{-1}$, which is about 15 -
20\% of the average velocity, while the full peculiar velocity of
the Local Group relative to the microwave background is $600\,
{\rm km}\,{\rm s}^{-1}$. The main inhomogeneity scales responsible
for such a high velocity are in the range 15 - 50 Mpc.

We can see that the standard theory of the formation of the
structure of the Universe faces no `mysterious' problems in
explaining the observed relative motion of matter in
quasi-homogeneous regions of the Universe ($\kappa < 1$). The
local flows are regular, smooth, and highly correlated. The
smallness of the random deviations in galaxy velocities from
average cold flows is explained with a profile of the initial
spatial density perturbation spectrum, contrarily to Chernin's
evolutionary influence of dark energy. At small distances, the
flows are quasi-Hubble (they are radial, the outflaw velocity
being in direct proportion to distance), but the Hubble parameter
depends on direction and on the observer's location. There is no
reason to modify the standard theory. The `Little Bang'
model\footnote{`The local cosmology' in Section 3.4 in [4], which
underlies the `Little Bang' model, is based on the limit of the
static gravitational field (see Eqns. (28)--(36) in [4]). This
contradicts the standard theory, where the gravitational potential
$\Phi (t, {\bf x})$ depends on time [see Eqn. (10)].} proposed in
Section 3 in [4] in order to explain the cold flows is beneath
criticism. In that model, the peculiar velocities of galaxies
`kicked' out of the Local Group must decrease as much as five
times under the gravitational influence of dark energy. As Fig. 1
shows, this will take more than 40 billion years.

\section{On `internal symmetry in cosmology'}

Chernin's suggestion to link relation (1) with mythological
`internal symmetry in cosmology' ([4], Section 5) causes serious
disagreement. In this respect, the crucial point for the author is
his introducing `Friedmann integrals.' To determine their values,
the author arbitrarily normalizes the scale factor (the value of
the parameter $R_0$ in Eqn. (57) in [4]) to the present horizon
size $R_0 = H_0^{\,-1}$. With the parameter $R_0$ chosen
differently, the equality of the `Friedmann integrals,' e.g.,
those for dark matter and dark energy, $A_{\rm D}$ and $A_{\rm
V}$, would be violated. It is clear that if $R_0 = H_0^{\,-1}$,
then the approximate relations between the `Friedmann integrals'
$$
A_{\rm B} \sim A_{\rm D} \sim A_{\rm R} \sim A_{\rm V} \eqno(15)
$$
are equivalent to the approximate relations
$$
\Omega _{\rm B} \sim \Omega _{\rm D} \sim \sqrt {\Omega _{\rm R}}
\sim {1\over \sqrt {\Omega _{\rm V}}} \eqno(16)
$$
for the present ratios of the energy densities to the critical
density. To verify this, it is sufficient to use the Friedmann
equation for the spatially flat or almost flat Universe at today's
instant, which gives
$$
A_\lambda = R_0 \bigl [ \Omega _\lambda (H_0 R_0)^2\bigr
]^{1/(1+3w_\lambda )}\,,
$$
for all sorts of matter $\lambda$, where $w_\lambda = p_\lambda
/\rho _\lambda$  (cf. Eqn. (59) in [4]). This last formula implies
the equivalence of relations (15) and (16) within an order of
magnitude. Because $\Omega _{\rm B}$, $\Omega _{\rm D}$, and
$\Omega _{\rm V}$ are presently quite close to unity [however, see
(17) and (18)], relation (16), in turn, is equivalent to the
approximate equality in (1) complemented with the same relation
for photons.

Thus, what the author thinks is a `symmetry' is, in fact, the
rephrased statement about the densities of different energy
components being close to each other. Introducing the `Friedmann
integrals' does not make relation (1) any clearer.

The author's approach to the flatness problem (`Dicke's problem')
is equally awkward. Without an extremely fine tuning of the
cosmological evolution initial data at the hot stage (the tuning
noted by Dicke, which is automatically fulfilled in the
inflationary theory), it is impossible to obtain small spatial
curvature in the present Universe. In the author's terms, this
means that, had it not been the fine-tuning of the initial data,
the dimensionless parameter in Eqn. (87) in [4] would have been
extremely high, and for the closed model, the equality between the
energy densities of matter and dark energy would have never been
satisfied. For example, for the closed Universe and the fixed
$\rho _{\rm V} \sim 10^{-29}\,{\rm g}\,{\rm cm}^{-3}$, it is
absolutely clear that if the curvature had contributed, e.g.,
$10^{-6}$ of the matter contribution to the Friedmann equation in
the nucleosynthesis period, the expansion of the Universe would
have changed to contraction and subsequent re-collapse long before
dark energy would play a role in the cosmological expansion.

In fact, relations like (1) or (16) involve some very interesting,
but still unclear issues. For example, the relation $\rho _{\rm D}
\approx \rho _{\rm B}$ being constant in time possibly indicates
the common origin of dark matter and the baryonic asymmetry of the
Universe. In spite of numerous attempts to account for this fact,
no satisfactory theoretical models have been proposed. The
relation $\rho _{\rm V} \approx \rho _{\rm D}$ is valid today but
includes quantities changing with time at different rates; this
suggests that the period of transition from the matter-dominated
to the dark-energy-dominated stage occupies a privileged position
on the time axis in terms of the structure (and hence life)
formation.

The fact of the existence of the large-scale structure is crucial
from the perspective of the coincidence problem. The relations
between the parameters $\Omega _{\rm R}$, $\Omega _{\rm V}$, and
$\Omega _{\rm M} =  \Omega _{\rm D} + \Omega _{\rm B}$,
$$
\Omega _{\rm R} \ll \Omega _{\rm M} \,, \qquad \Omega _{\rm V}
\,{}^<_\sim \,\Omega _{\rm M}\,,        \eqno(17)
$$
have a direct impact on the possibility of the generation of the
structure of the Universe because gravitational instability is not
realized at the radiation- and dark-energy-dominated stages and
develops only if nonrelativistic matter dominates. But for the
structure to be formed, one more condition must be satisfied: the
initial perturbation amplitude must be just right to fit the
`window' of gravitational instability, thus giving rise to
inhomogeneities. In our Universe, the two necessary conditions are
satisfied: the initial perturbations (of the order $10^{-5}$)
manage to grow and form the large-scale structure of the Universe
during the time `window' from 0.3 to 20 billion years. The
condition
$$
\Omega _{\rm R} \ll \Omega _{\rm B} \, {}^<_\sim \, \Omega _{\rm
D}\,, \eqno(18)
$$
in turn, is necessary for forming stars in nonlinear halos of dark
matter.

Any detailed discussion of these and similar issues would require
a new review (in this respect, see, e.g., [1, 3, 19--21]). Here,
we find it important to emphasize that the outlined range of
issues is much richer than it may seem after reading Chernin's
article.

\section{On the physical nature of dark energy}

Almost everywhere in [4], Chernin identifies dark energy with the
vacuum energy, while other possibilities are mentioned just
briefly. We want to point out that a no less attractive point of
view is to relate dark energy to a new superweak and superlight
field, which can be a quintessence, a phantom field, etc.

It is appealing because, among other things, it is very hard to
explain the vacuum energy value, which is nonzero and still
extremely small compared to the energy scales of the known
interactions (see Section 1). It is much easier to imagine the
vacuum energy relaxing practically to zero at some stage of the
evolution of the Universe (long before the known stages). There
are examples of such mechanisms in the literature [22, 23]. In
this framework, it is natural for dark energy to be the energy of
a new field rather than the vacuum energy. Another heuristic
argument is that the present stage of the accelerated expansion of
the Universe looks qualitatively similar to the inflation stage
and differs from the latter `only' in the energy density value
and, hence, the Hubble parameter. Dark energy in the form of a
superweak field could be a less energetic counterpart of the
inflaton, a field commonly used in inflationary theories.

If dark energy is the energy of a new field, the parameter $w_{\rm
V}$ that relates pressure and energy density in accordance with
the equation $p_{\rm V}=w_{\rm V} \rho _{\rm V}$, differs from -1
(by the way, it does not have to be constant in time), and the
dark energy density depends on time. However, we emphasize that in
most models of this type, the parameter $w_{\rm V}$ is
automatically close to the vacuum value $w=-1$, and hence the
observational limit $|w_{\rm V}+1| < 0.1$ hardly constrains the
existing models yet.

Finally, we note that the accelerated expansion may be caused by
gravitation theory modified at superlarge scales and cosmological
times. One of the possibilities is here related to the extra
spatial dimensions of infinite size (for example, see Ref. [24]),
although attempts to construct such models have experienced
internal contradictions so far. Another possibility, more
realistic from the standpoint of theoretical realization, is the
extension of General Relativity to the scalar--tensor theory of
gravity [25, 26].

Thus, the Universe's accelerated expansion may be the first
evidence of new physical phenomena occurring at cosmological and
maybe other scales. Various models of the accelerated expansion
differ in the dark energy density dependence on time. The search
for this dependence and its detailed study are important problems
of observational cosmology, which must eventually allow revealing
the physical nature of dark energy.

\section{Additional comments}

Chernin's article ought to be read with caution. For example, not
all researchers are that enthusiastic about the extravagant model
by Luminet et al.; peculiarities, if any, in the angular spectrum
of the cosmic microwave background can be explained in a less
exotic way. Almost the same applies to the model by Arkani-Hamed,
Dimopoulos, and Dvali (known as the ADD model). This model
definitely played a great role in presenting the idea that extra
spatial dimensions can be large (or even infinitely large), but it
is unlikely that nature follows this way. We note further that
Chernin's paper is replete with terms that are not commonly
accepted, such as EG'vacuum, Q-vacuum, and Friedmann integrals.

To summarize, we recommend that the concerned reader form a
reasoned opinion on issues mentioned in Chernin's article using
alternative reviews on this topic, e.g., Refs. [20, 27--30].

\section{Conclusion}

The discovery of dark energy dotted the $i\,$'s and crossed the
$t\,$'s in observational cosmology. The standard cosmological
model ($\Lambda CDM$) fitting the whole set of observational data
arose for the first time in the development of science. Nowadays,
it has no serious rivals. The standard model describes both the
evolution of the Universe as a whole and the generation of its
structure remarkably well. In spite of the influence of dark
energy, the structure is still being generated and this will
continue for another ten billion years or so.

At the same time, with dark energy having been recognized to
exist, the situation in physics has dramatically changed and we
see our knowledge of the microworld as incomplete. It is a safe
bet to say that revealing the physical nature of dark energy is
the central problem of natural science.

\section*{Acknowledgments} The authors are grateful to A.G. Doroshkevich,
M.B. Libanov, E.V. Mikheeva, and V.N. Strokov for the useful
discussions. The work was supported in part by the Russian
Foundation for Basic Research, Grants 07-02-00886 and 08-02-00473.

\end{document}